\documentclass[12pt,preprint]{aastex}
\usepackage{psfig}

\newcommand\mzon   {M$_{\odot}$}
\newcommand\pp     {$\pm$}

\newcommand\Lunit   {erg s$^{-1}$}
\newcommand\funit   {erg cm$^{-2}$ s$^{-1}$}

\slugcomment{Accepted for publication in ApJ Letters: March 19, 2004}

\begin{document}

\title{Monitoring {\itshape Chandra} observations of the 
quasi-persistent neutron-star X-ray transient MXB 1659--29 in
quiescence: the cooling curve of the heated neutron-star crust}

\author{Rudy Wijnands\altaffilmark{1,2}, Jeroen Homan\altaffilmark{3},
Jon M. Miller\altaffilmark{4,5}, Walter H. G. Lewin\altaffilmark{3}}

\altaffiltext{1}{School of Physics and Astronomy, 
University of St Andrews, North Haugh, St Andrews, Fife KY16 9SS,
Scotland, UK; radw@st-andrews.ac.uk}

\altaffiltext{2}{Present address: Astronomical Institute ``Anton Pannekoek'',
University of Amsterdam, Kruislaan 403, 1098 SJ, Amsterdam, the
Netherlands; rudy@science.uva.nl}

\altaffiltext{3}{Center for Space Research, Massachusetts Institute of
Technology, 77 Massachusetts Avenue, Cambridge, MA 02139, USA;
jeroen@space.mit.edu, lewin@space.mit.edu}

\altaffiltext{4}{Harvard-Smithsonian Center for Astrophysics, 
60 Garden Street, Cambridge, MA 02139, USA; jmmiller@head.cfa.harvard.edu}

\altaffiltext{5}{NSF Astronomy \& Astrophysics Fellow}

\begin{abstract}

We have observed the quasi-persistent neutron-star X-ray transient and
eclipsing binary MXB 1659--29 in quiescence on three occasions with
{\it Chandra}. The purpose of our observations was to monitor the
quiescent behavior of the source after its last prolonged ($\sim$2.5
years) outburst which ended in September 2001.  The X-ray spectra of
the source are consistent with thermal radiation from the neutron-star
surface. We found that the bolometric flux of the source decreased by
a factor of 7--9 over the time-span of 1.5 years between our first and
last {\it Chandra} observations. The effective temperature also
decreased, by a factor of 1.6--1.7. The decrease in time of the
bolometric flux and effective temperature can be described using
exponential decay functions, with $e$-folding times of $\sim$0.7 and
$\sim$3 years, respectively. Our results are consistent with the
hypothesis that we observed a cooling neutron-star crust which was
heated considerably during the prolonged accretion event and which is
still out of thermal equilibrium with the neutron-star core. We could
only determine upper-limits for any luminosity contribution due to the
thermal state of the neutron-star core. The rapid cooling of the
neutron-star crust implies that it has a large thermal
conductivity. Our results also suggest that enhanced cooling processes
are present in the neutron-star core.

\end{abstract}


\section{Introduction}

Neutron stars in low-mass X-ray binaries accrete matter from solar
mass companions. Among those systems, the sub-group of neutron-star
transients spend most of their time in quiescence during which hardly
any or no accretion occurs. However, these transients sporadically
become very bright ($>$$10^{36-38}$~\Lunit) owing to a huge increase
in the accretion rate onto their neutron stars. During those
outbursts, these sources can be readily studied with the available
X-ray instruments, but obtaining high quality quiescent data remains a
challenge. In spite of this, several systems have now been studied in
detail: they typically exhibit 0.5--10 keV luminosities of
$10^{32-33}$~\Lunit~and their spectra are usually dominated by a soft
component which can be described by a thermal model. This emission is
thought to be due to the cooling of the neutron star which has been
heated during the outbursts (Brown, Bildsten, \& Rutledge 1998;
Campana et al.~1998a).

Most neutron-star transients are active for only weeks to months, but
several systems have remained active for years and even decades (the
'quasi-persistent' neutron-star transients; Wijnands et
al. 2003). Wijnands et al.~(2001) realized that those systems are
excellent targets to study the effects of accretion on the behavior of
neutron stars by observing them in quiescence.  The accreting material
is expected to have a larger effect on the neutron stars in such
systems than on the neutron stars in short-duration transients
(Wijnands et al.~2001; Rutledge et al.~2002). In the latter systems,
the crust is only marginally heated during the outbursts and will
quickly return to thermal equilibrium with the core after the end of
the outbursts. In the quasi-persistent transients, however, the crust
is heated to high temperatures and becomes significantly out of
thermal equilibrium with the core (Rutledge et al.~2002). After the
end of the prolonged outbursts, it will cool until it returns to
equilibrium with the core.  The exact cooling time depends on the
thermal conductivity of the crust, the core cooling processes, and the
accretion history of the source.

KS 1731--260 was the first quasi-persistent transient to be studied in
detail in quiescence.  It was observed using {\it Chandra} shortly
after the end of its $\sim$12.5 year outburst (Wijnands et al.~2001)
and it was found to have a luminosity of $\sim10^{33}$~\Lunit~(for a
distance $d=7$ kpc; 0.5--10 keV). Half a year later it was observed
with {\it XMM-Newton} and it was found that its luminosity had
decreased by a factor of 2--3 (Wijnands et al.~2002b). Using the
cooling curves calculated by Rutledge et al.~(2002), this drop in
brightness can be explained if the neutron star has a large crustal
conductivity and enhanced core cooling processes. In September 2001, a
second quasi-persistent neutron-star transient (MXB 1659--29) turned
off after having accreted for $\sim$2.5 years.  Wijnands et al.~(2003)
obtained a {\it Chandra} observation of this source within a month
after the end of its outburst and detected it at a luminosity of
$\sim3-4 \times 10^{33}$~\Lunit~(0.5--10 keV; $d=10$ kpc). Several
years before this outburst, the source was observed with {\it ROSAT},
but could not be detected (Verbunt 2001). The flux upper limit was
$\sim$10 times lower than the {\it Chandra} flux (Oosterbroek et
al.~2001; Wijnands 2002). Wijnands et al.~(2003) concluded that during
the {\it Chandra} observation the observed radiation was due to a hot
crust and not associated with the core.

\section{Observations, analysis, and results}

{\it Chandra} observed MXB 1659--29 twice for $\sim$27 ksec: on
October 15, 2002 (the 2002 observation), and on May 9, 2003 (the 2003
observation).  We also used the $\sim$19 ksec observation performed on
October 15--16, 2001 (the 2001 observation; Wijnands et
al.~2003). During all observations the ACIS-S3 chip was used. The data
were reduced and analyzed using CIAO 3.0.  To make use of the latest
calibration products, we reprocessed the 2001 observation. A minor
background flare occurred during the 2003 observation (factor of
$\sim$2; lasting $\sim$2 ksec). Its effect on the quality of the
source data was negligible and we did not to remove this flare from
the data. No flares occurred during the other observations.

For each observation, we extracted the number of source photons, the
light curve, and the spectrum, using a circle with a radius of 3$''$
as source extraction region and an annulus with an inner radius of
$7''$ and an outer radius of 22$''$ as background region. We detected
948\pp31, 263\pp16, and 107\pp10 counts (0.3--7 keV; background
corrected) for the 2001, 2002, and 2003 observations, respectively,
resulting in corresponding count rates of 0.050\pp0.002,
0.0097\pp0.0006, and 0.0039\pp0.0004 counts~s$^{-1}$. Wijnands et
al. (2003) observed an eclipse and dipping behavior during the 2001
observation (similar to the outburst behavior of the source; Lewin
1979; Cominsky et al. 1983; Cominsky \& Wood 1984, 1989). To search
for eclipsing behavior during the 2002 and 2003 observations, we
determined the orbital phase range covered by those observations using
the time of the eclipse in the 2001 observation as the reference
time. Given the orbital phase range traced during each observation, we
expect to see a single eclipse per observation and, as anticipated, we
did not detect any photons during the expected eclipse intervals.
However, we also found that no photons were detected during several
time intervals (of equal duration as the lengths of the eclipses) at
different phases of the orbital period. Therefore, without prior
knowledge of the eclipsing nature of MXB 1659--29, we could not have
concluded that we saw eclipses during the 2002 and 2003
observations. Owing to the limited statistics of the 2002 and 2003
observations, no conclusions can be drawn about possible dipping
behavior during these observations.

When extracting the spectra, we used all data, including those taken
during the intervals of eclipses and possible dipping behavior. The
eclipses could not be removed from the data before extracting the
spectra because the uncertainties in the ephemeris presented by
Oosterbroek et al. (2001) are sufficiently large so that the exact
start and end times of the expected eclipses could not be
determined. Instead we decreased the exposure time in the resulting
spectral files by 900 s since the eclipse duration during outburst was
found to be $\sim$900 s (Wachter et al. 2000) and Wijnands et
al. (2003) reported an eclipse duration of 842\pp90 seconds for the
2001 observation. Small differences in the eclipse duration might be
present between the observations but the expected effects on the
resulting fluxes will be marginal. We also did not remove the data
obtained during the dipping interval observed in the 2001
observation. Such dipping intervals are likely present during the
other two observations but they cannot be identified in the light
curves due to limited statistics. For those two observations all data
had to be used and to obtain a homogeneous data selection across
observations, we included the dipping interval observed during the
2001 observation. Wijnands et al. (2003) found evidence that this
dipping behavior is likely due to a change in internal absorption in
the system and not due to actual changes in the neutron-star
properties. Therefore, the inclusion of the (possible) dipping
intervals will likely result in a somewhat higher column density
($N_{\rm H}$) in the spectral fits than the true interstellar $N_{\rm
H}$ toward the source, but should not significantly impact other
source properties.

We grouped the spectra in bins of 15 counts to validate the use of the
$\chi^2$ fitting method and simultaneously fitted the three spectra
using Xspec (Arnaud 1996). A variety of one-component
models\footnote{E.g., a power-law model could fit the spectra but with
an index of 4.7--5.8 suggesting soft thermal spectra. We also fitted a
NSA plus power-law model to determine the upper-limits on the
contribution of such a power-law tail to the 0.5--10 keV flux. Those
limits are $<$20\%--25\%, $<$35\%--45\%, and $<$50\%--100\%, for the
2001, 2002, and 2003 observation, respectively. The range of upper
limits is due to the range assumed in photon indices (between 1 and
2).} could fit the individual spectra satisfactorily, but since we
expect that the X-rays from MXB 1659--29 are due to the cooling of the
neutron-star surface, for this paper we only fit the data using a
neutron-star hydrogen atmosphere model (NSA; for weakly magnetized
neutron stars; Zavlin et al. 1996). In such models the normalization
is given by $1/d^2$, with $d$ in pc. The distance should be constant
between observations and therefore we left the normalization tied
among the different spectra (when leaving the normalizations free
between observations, we find that they are consistent with each
other). We expect the $N_{\rm H}$ toward the source to be very similar
between observations (only minor variations are expected due to
variable internal absorption) and this parameter was also tied. We
assume a 'canonical' neutron star with a radius of 10 km and a mass of
1.4 \mzon.

From the fits, we found that the normalization was
$1.4^{+2.2}_{-0.8}\times 10^{-8}$ which yields a source distance of
5--13 kpc. This is consistent with the distance range given in the
literature (10--13 kpc; Oosterbroek et al. 2001; Muno et
al. 2001). However, we found that the errors on the fit parameters
were dominated by the large uncertainties in the normalization and did
not allow us to realize the full potential of the data. If the source
distance were established through an independent method, we could fix
the normalization in the NSA models, resulting in considerably smaller
errors on the remaining fit parameters. Therefore, instead of leaving
the normalization as a free parameter, we fixed it so that it
corresponded to a distance of 5, 10, and 13 kpc, covering the full
range of allowed distances obtained when the normalization was a free
parameter.  To estimate the bolometric fluxes ($F_{\rm bol}$) we
extrapolated the model to the energy range 0.01--100 keV which gives
approximate bolometric fluxes\footnote{We verified that the 0.01--100
keV fluxes approximate $F_{\rm bol}$ by calculating the bolometric
luminosity $L_{\rm bol} = 4\pi\sigma R^{2}_{\infty} T^{\infty 4}_{\rm
eff}$, with $\sigma$ Stefan-Boltzmann constant, $T^{\infty}_{\rm eff}$
the effective temperature (at infinity), and $R_{\infty}$ the
neutron-star radius (at infinity). The 0.01--100 keV fluxes were
indeed consistent with the calculated $F_{\rm bol}$. We use the
measured fluxes because their errors takes into account the
uncertainties in $N_{\rm H}$ and the $T^{\infty}_{\rm eff}$ obtained
for all observations. The $L_{\rm bol}$ errors are only calculated
using the $T^{\infty}_{\rm eff}$ errors during one specific
observation. \label{footnote}}.  To calculate the flux errors, we
fixed each free fit parameter (only one at a time) either to its
minimum or maximum allowed value. After that we refitted the data and
recalculated the fluxes. This process was repeated for each free
parameter and the final flux range determined the flux errors.  The
fit parameters obtained are listed in Table~\ref{tab:spectra}.

\begin{figure}
\begin{center}
\begin{tabular}{c}
\psfig{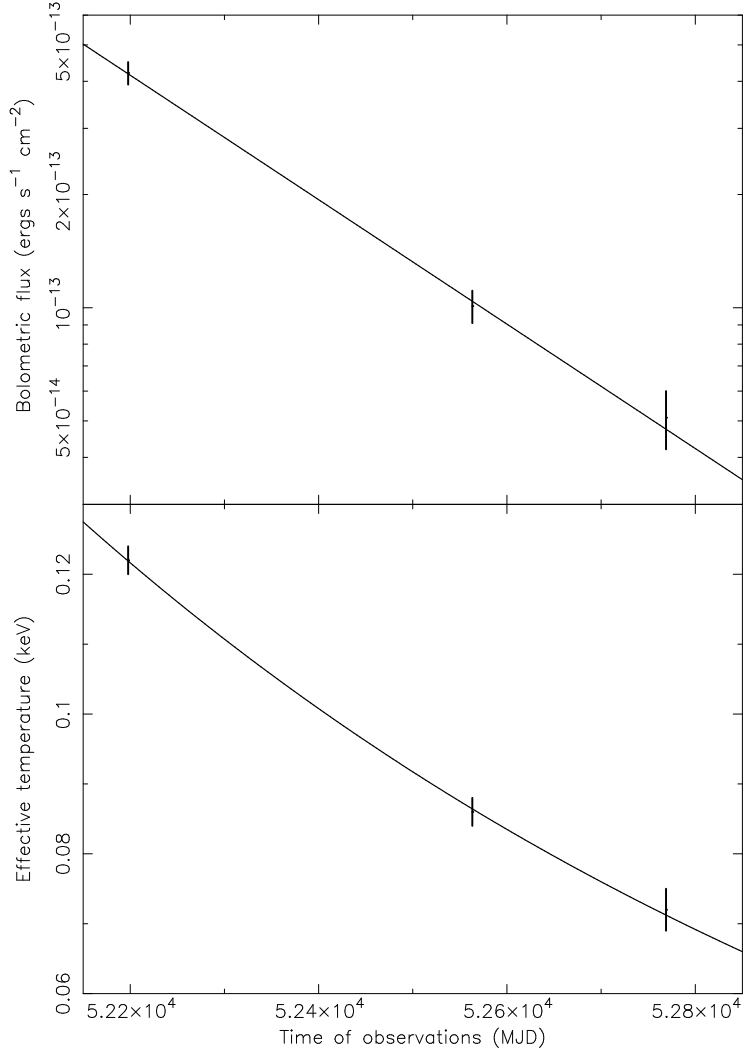}
\end{tabular}
\figcaption{
The bolometric flux (top panel) and effective temperature (bottom
panel; for an observer at infinity) of the neutron-star crust as a
function of time (as obtained with the neutron-star hydrogen
atmosphere model for weakly magnetized neutron stars of Zavlin et
al. 1996). The solid curves are the best fit exponential function
through the data points.  The bolometric fluxes are plotted on a
logarithmic scale, but for clarity, the effective temperatures are
plotted on a linear scale. The data used in this figure are those
obtained when assuming a distance of 10 kpc in the spectral fits; the
figures for a distance of 5 or 13 kpc are very similar and therefore
we omit them. Only the absolute values of the bolometric flux and
effective temperature are different but the overall decay trend is
nearly identical.
\label{fig:decay} }
\end{center}
\end{figure}

This table shows that $T^{\infty}_{\rm eff}$ and $F_{\rm bol}$
decreased in time (Fig.~\ref{fig:decay}). We fitted the
$T^{\infty}_{\rm eff}$ and $F_{\rm bol}$ curves with an exponential
decay function $y(t) = c_0 e^{- {t - t_0\over
\tau}}$, with $c_0$ a normalization constant, $t_0$ the start time,
and $\tau$ the $e$-folding time. We found that the other fit
parameters were not very sensitive to the value of $t_0$, but when
$t_0$ was left free it had adverse effects on the errors on those
parameters.  Therefore, we fixed $t_0$ to MJD 52159.5 which
corresponds to midday September 7, 2001 (the last day MXB 1659--29 was
found to be active; Wijnands et al. 2002a) and which can be regarded
as an approximation of the time when $T^{\infty}_{\rm eff}$ and
$F_{\rm bol}$ began to decrease.  The assumed exponential functions
could adequately describe the decrease in $T^{\infty}_{\rm eff}$ and
$F_{\rm bol}$ (Fig.~\ref{fig:decay}; alternative functions did not
provide adequate fits).  We found that $\tau$ and $c_0$ for the
$F_{\rm bol}$ curve were 289\pp37, 262\pp33, and 254\pp29 days, and
70\pp9, 48\pp6, and 43\pp6 $\times 10^{-14}$ \funit, when assuming a
distance of 5, 10, or 13 kpc, respectively, in the spectral fits.  The
corresponding $\tau$ and $c_0$ for the $T^{\infty}_{\rm eff}$ curve
were 1153\pp160, 1060\pp126, and 1055\pp112 days, and 0.099\pp0.004,
0.126\pp0.004, and 0.139\pp0.004 keV. We saw no evidence that the
curves approached a rock-bottom value: we found an upper limit on such
a value of 3.5--7.5 $\times 10^{-14}$ \funit~for the $F_{\rm bol}$
curve (resulting in bolometric luminosity limits of 2.2--7.0
$\times 10^{32}$ \Lunit), and 0.06--0.07 keV for the $T^{\infty}_{\rm
eff}$ curve.
 
\section{Discussion}

We have presented monitoring {\it Chandra} observations of MXB
1659--29 in quiescence. The first observation was taken only a month
after the end of its last outburst which lasted 2.5 years; the second
and third observations were taken $\sim$1 and $\sim$1.5 years after
this initial one. Because it is expected that the emission should be
dominated by thermal emission from the hot neutron-star crust (see
Wijnands et al. 2003), we fitted the data with a NSA model for weakly
($B<10^{8-9}$ G) magnetized neutron stars. We found that $F_{\rm bol}$
decreased by a factor of $\sim$8 in $\sim$1.5 years and the rate of
decrease followed an exponential decay function.  Furthermore, $T_{\rm
eff}^{\infty}$ also decreased and the rate of decrease again followed
an exponential decay function. We found that the $e$-folding time of
the $T_{\rm eff}^{\infty}$ curve was consistent with four times that
of the $F_{\rm bol}$ curve, as expected if the emission is caused by a
cooling black body for which the bolometric luminosity is given by
$L_{\rm bol}=4\pi\sigma R^{2}_{\infty} T^{\infty 4}_{\rm eff}$ (see
footnote~\ref{footnote}): if $T^{\infty}_{\rm eff}$ decays
exponentially, $L_{\rm bol}$ (and thus $F_{\rm bol}$) will also decay
exponentially but with an $e$-folding time four times smaller than
that of $T^{\infty}_{\rm eff}$, exactly what we observe.

Our results support the suggestion that the crust was heated to high
temperatures during the prolonged accretion event, which ended a month
before our first observation, and that it is now cooling until it
reaches thermal equilibrium with the core.  Rutledge et al. (2002)
calculated cooling curves for the neutron star in KS 1731--260,
assuming different behaviors of the crustal micro-physics and the core
cooling processes. Those curves can be used as a starting point to
investigate how our results of MXB 1659--29 could be explained. Of
those curves, only the one which assumes a large crustal conductivity
and the presence of enhanced core cooling processes exhibits a large
luminosity decrease in the first two years after the end of the last
outburst, suggesting that the neutron star in MXB 1659--29 has similar
properties.  This conclusion was already tentatively reached by
Wijnands et al. (2003) based on a comparison of the luminosity seen
during the October 2001 {\it Chandra} observation with the
significantly lower luminosity upper-limit found with {\it ROSAT}. But
detailed cooling curves for the neutron star in MXB 1659--29 need to
be calculated to fully explore (and exploit) the impact of our
observations on our understanding of the structure of neutron
stars. The cooling curves calculated by Rutledge et al. (2002) for KS
1731--260 only give us a hint of the behavior of MXB 1659--29 because
they depend on the long-term ($>10^4$ years) accretion history of the
source. For KS 1731--260, this long-term accretion behavior was quite
unconstrained due to large uncertainties in the averaged duration of
the outbursts, the time-averaged accretion rate during the outbursts,
and the time the source spent in quiescence. However, the accretion
history of MXB 1659--29 over the last three decades is much better
constrained (Wijnands et al. 2003), which will help to reduce the
uncertainties in its long-term averaged accretion history allowing for
more detailed cooling curves to be calculated for MXB 1659--29. This
might help to constrain the physics of the crust better for MXB
1659--29 than for KS 1731--260. The only significant uncertainty left
is that of the source distance; however, we found that this only
affects the exact values of the bolometric fluxes and the effective
temperatures, but not their rate of decay.

Our 0.5--10 keV flux during the May 2003 {\it Chandra} observation is
still higher than the upper limit found with {\it ROSAT}, suggesting
that the crust will cool even further in quiescence and that we have
not yet reached thermal equilibrium between the crust and
core. Further monitoring observations are needed to follow the cooling
curve of the crust to determine the moment when the crust is thermally
relaxed again.  When this occurs, no significant further decrease of
the quiescent luminosity is expected and from this bottom level the
state of the core can be inferred. As of yet, we have found no
evidence that the flux and temperature are reaching a leveling-off
value, associated with the temperature of the core, although the
limits we obtained are not very stringent.

Jonker, Wijnands, \& van der Klis (2004) suggested that the difference
in luminosity of MXB 1659--29 between the {\it ROSAT} non-detection
and the 2001 {\it Chandra} observation might be due to differences in
residual accretion rate onto the surface. Residual accretion could
indeed produce soft spectra (e.g., Zampieri et al. 1995), but to
explain the exponential decay we observe for $F_{\rm bol}$ and $T_{\rm
eff}^{\infty}$, the residual accretion rate must also decrease
exponentially with a timescale of a year. Although this cannot be
completely ruled out, we believe this is unlikely since other
neutron-star transients have been observed to reach their quiescent
states on timescales of only tens to several tens of days at the end
of their outbursts (e.g., Campana et al. 1998b; Jonker et al. 2003)
and the variations in accretion rate tend to be more
stochastic. Moreover, if the neutron star has a significant magnetic
field strength, this might inhibit material from reaching the surface
when accreting at the inferred low rates.

\clearpage
\begin{figure}
\begin{center}
\begin{tabular}{c}
\psfig{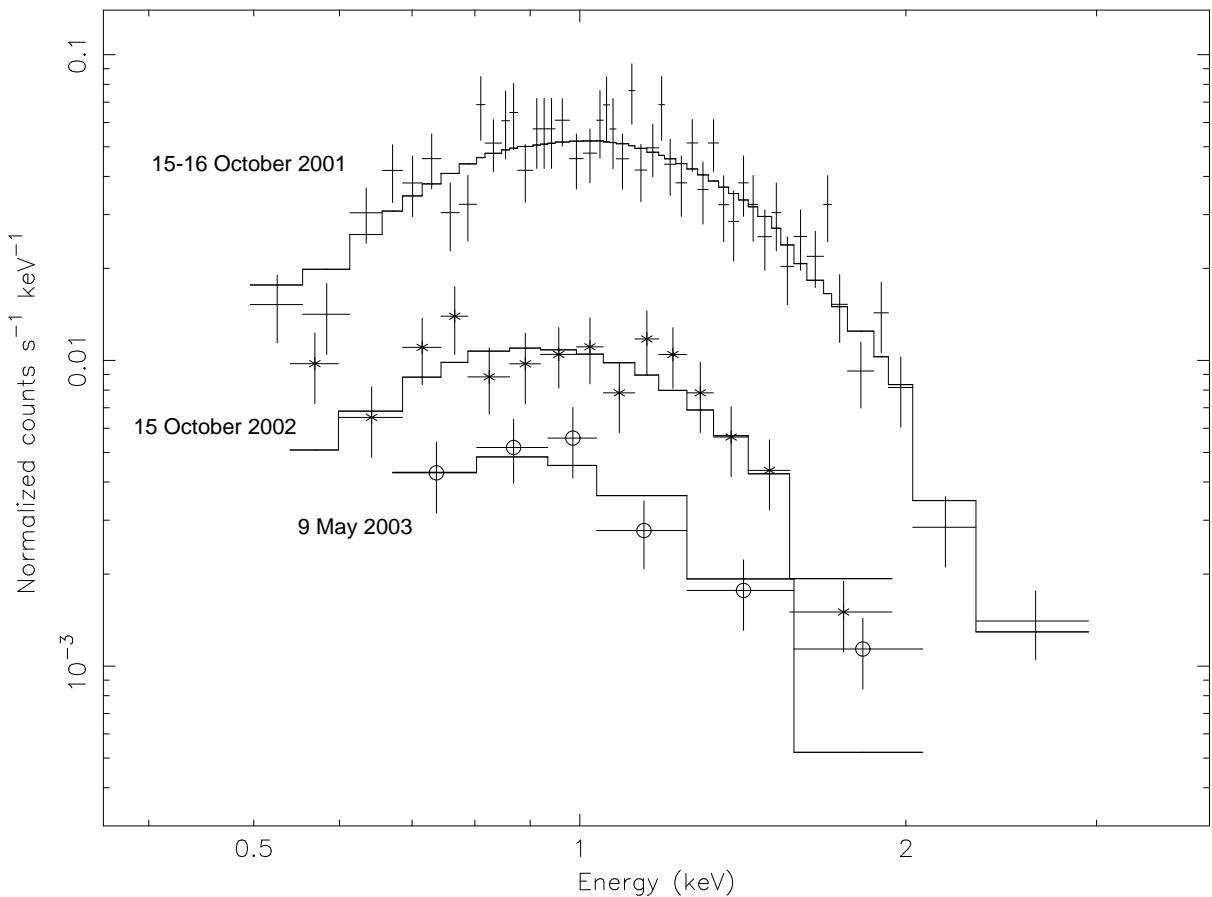}
\end{tabular}
\figcaption{
The {\it Chandra} X-ray spectra obtained during the quiescent state of
MXB 1659--29. The top spectrum was obtained on October 15--16, 2001,
the middle spectrum (indicated by the crosses) was obtained on October
15, 2002, and the bottom spectrum (indicated by the open circles) was
obtained on May 9, 2003. The solid lines through the spectra indicate
the best fit neutron-star hydrogen atmosphere model (that of Zavlin et
al. 1996; for weakly magnetized neutron stars).
\label{fig:spectra} }
\end{center}
\end{figure}

\clearpage

\clearpage

\begin{deluxetable}{lccc}
\tablecolumns{4}
\tablewidth{0pt} 
\tablecaption{Spectral results for MXB 1659--29\label{tab:spectra}}
\tablehead{
Parameter                         &\multicolumn{3}{c}{Distance assumed} \\ 
                                  & 5 kpc                & 10 kpc               & 13 kpc  }
\startdata  
$N_{\rm H}$ ($10^{21}$ cm$^{-2}$) & 2.8\pp0.3            & 1.8\pp0.2            & 1.5\pp0.2                    \\
\hline
\multicolumn{4}{l}{$kT_{\rm eff}^{\infty}$ (keV)} \\
~~~2001                           & 0.096\pp0.002        & 0.122\pp0.002        & 0.134\pp0.002                \\
~~~2002                           & 0.069\pp0.002        & 0.086\pp0.002        & 0.094$^{+0.002}_{-0.003}$    \\
~~~2003                           & 0.059\pp0.002        & 0.072\pp0.003        & 0.079\pp0.003                \\
\hline
\multicolumn{4}{l}{Flux ($10^{-14}$ \funit; 0.5--10 keV; unabsorbed)}\\
~~~2001                           & 41.8\pp3.2           & 31.3\pp2.3           & 28.4\pp2.1                       \\
~~~2002                           &  9.1\pp1.0           &  6.4$^{+0.8}_{-0.6}$ & 5.7\pp0.6                    \\
~~~2003                           &  4.0$^{+0.8}_{-0.2}$ &  2.8$^{+0.7}_{-0.5}$ & 2.5\pp0.4                    \\
\hline
\multicolumn{4}{l}{Bolometric flux  ($10^{-14}$ \funit; unabsorbed)}\\
~~~2001                           & 61.6\pp4.2           & 42.1\pp2.9           & 37.7\pp2.8                       \\
~~~2002                           & 16.9\pp1.5           & 10.1\pp1.0           & 8.5\pp0.8                    \\
~~~2003                           &  8.9$^{+1.4}_{-0.4}$ & 5.1\pp0.9            & 4.2\pp0.6                    \\
\hline
$\chi^2$/d.o.f.                   & 59.2/65              & 56.1/65              & 58.6/65                      \\    
\enddata

\tablenotetext{\,}{Note: The error bars represent 90\% confidence
levels. We used a neutron-star mass of 1.4 \mzon~and radius of 10 km
and the neutron-star hydrogen atmosphere model for weakly magnetized
neutron stars of Zavlin et al. 1996. }

\end{deluxetable}

\end{document}